\documentclass[aps,prd,twocolumn,preprintnumbers,amsmath,amssymb,
showpacs,floatfix]{revtex4}

\usepackage{graphicx}
\usepackage{dcolumn}
\usepackage{bm}
\usepackage{epsfig}
\usepackage{graphics}
\usepackage{latexsym}

\begin{document}

\preprint{}

\title{ Hybrid Inflation in Quasi-minimal Supergravity with Monotonic Inflationary Potential}

\author{C. Panagiotakopoulos}
\email{costapan@eng.auth.gr}
\affiliation{ School of Rural and Surveying Engineering, Faculty of 
Engineering, Aristotle University of Thessaloniki, Thessaloniki 
54124, Greece}


\begin{abstract}
We show how supersymmetric hybrid inflation with scalar spectral index
$n_{\rm s}\simeq 0.96 - 0.97$ is realized in the context of quasi-minimal supergravity
if we insist that the inflationary potential not exhibit any local minima.
We also address the problem of the initial conditions for both monotonic
and non-monotonic inflationary potentials.
\end{abstract}


\maketitle

\section{Introduction}

The simplest supersymmetric (SUSY) hybrid 
inflation model \cite{dvali} with weak radiative corrections and minimal
 K\"{a}hler potential predicts a scalar spectral index $n_{\rm s}$
which  lies close to 0.98. An advantage of this scenario is that inflation
is realized for inflaton values well below the reduced Planck scale
$m_{\rm P} \simeq 2.44\times 10^{18}~{\rm GeV}$ (which is set
to 1 throughout the rest of the paper) and is
 insensitive to supergravity corrections due to a cancellation
of the supergravity-induced inflaton mass squared term in such
a minimal model \cite{copeland}.
Inclusion of the first correction involving the inflaton field in the  K\"{a}hler
potential destroys the cancellation and generates an inflaton mass squared
which seriously affects the inflationary scenario even at weak coupling
of the inflaton. Assuming that the mass squared of the inflaton is positive
\cite{pana0} in order to avoid the generation of local minima in the
inflationary potential the value of the spectral index increases and soon
the spectrum of density perturbations turns from red to blue.
If, instead, we overlook the danger of the inflaton being trapped in a 
local minimum and allow for a negative inflaton mass squared \cite{bastero}
we may lower the value of the spectral index to lie in the presently
favored range $n_{\rm s}\simeq0.96-0.97$  \cite{planckcosm}.

Our purpose here is to explore in detail the range of the parameters
for which the inflationary potential is with certainty monotonic
 in spite of the inflaton mass squared being negative and, of course, the
characteristics of ``observable"  inflation are acceptable. The existence
of a region in the parameter space where the local minimum of the
inflationary potential disappears was known to the authors of \cite{bastero}
but apparently they decided to concentrate on the region exhiditing local
extrema and examine the inflationary scenario taking place for inflaton
field values smaller in size than the position of the local maximum.
The region where the potential is monotonic is also qualitatively
described in the numerical  exploration of \cite{rehman0} but no
particular attention is payed to it.

We also address the problem of the initial conditions \cite{tet} of such
hybrid inflation scenarios by invoking an early inflationary stage
\cite{initial, pana1, pana2} considering inflationary potentials which are
either monotonic or non-monotonic. Obviously, the existence of a local
minimum in the inflationary potential complicates this already difficult problem.

The structure of the paper is the following. In Sec. \ref{infl} we present
the SUSY hybrid inflation model and determine the range of the parameters
for which the inflationary potential is monotonic. Sec. \ref{ini} is devoted to
the initial condition problem. Finally Sec. \ref{concl} contains our conclusions.

\section{The SUSY Hybrid Inflation Model}
\label{infl}

For definiteness, we consider a SUSY model based on the 
left-right symmetric gauge group $G_{\rm{LR}}=SU(3)_{\rm c}
\times SU(2)_{\rm L}\times SU(2)_{\rm R}\times U(1)_{B-L}$. 
The superfields of the model 
which are relevant for inflation are a gauge 
singlet $S$ and a conjugate pair of Higgs superfields 
$\Phi$ and $\bar{\Phi}$ belonging to the $(1,1,2)_{1}$ and 
$(1,1,2)_{-1}$ representations of $G_{\rm LR}$, respectively.
Here the subscripts denote $U(1)_{B-L}$ charges.  
The fields  $\Phi$ and $\bar{\Phi}$ acquire  vacuum expectation values (VEVs) 
which break $G_{\rm LR}$ to the standard model gauge group $G_{\rm SM}$.
In addition we impose a global  $U(1)$ R symmetry under which $S$ and
the superpotential have charge 1 with $\Phi$ and $\bar{\Phi}$ being neutral.

The superpotential component which is relevant for our discussion is
\begin{equation}
W=\kappa S\left(M^2-\Phi{\bar \Phi}\right)
\label{W}
\end{equation}
with the K\"{a}hler potential taken to be
\begin{eqnarray}
\label{kaehler}
K=|S|^2+|\Phi|^2+|\bar \Phi|^2 +\frac{\alpha}{4}|S|^4.
\end{eqnarray}
Here $M$ is a superheavy mass and $\kappa$, $\alpha$
 are dimensionless constants which are all assumed to be real and positive. 

The F-term potential turns out to be
\begin{equation}
\label{pot1}
V_F=\kappa^2 V_0\exp{K}
\end{equation}
with
\begin{eqnarray}
V_0&=&|M^2-\Phi \bar \Phi|^2\left(\left(1+|S|^2+\frac{\alpha}{2}|S|^4\right)^2\frac{1}{1+\alpha|S|^2}\right.\nonumber \\
&&\quad\quad\quad\quad\quad\quad\left.-3|S|^2+|S|^2\left(|\Phi|^2+|\bar \Phi|^2\right){\frac{}{}}\right)\nonumber\\
&&+|S|^2\left(|\Phi|^2+|\bar \Phi|^2+4|\Phi|^2|\bar \Phi|^2\right)\nonumber\\
&&-2 M^2|S|^2\left(\Phi\bar \Phi+H.c.\right)
\end{eqnarray}
The SUSY minimum of the potential, $S=0, |\Phi|=|\bar \Phi|=M$  lies along
the D-flat direction $\Phi=\bar \Phi^*$. 
For $|S|>|S_{\rm c}|\simeq M$, where $S_{\rm c}$ is a critical value of $S$, 
the masses squared of $\Phi$, $\bar \Phi$
are positive and as a consequence the choice $\Phi=\bar \Phi=0$ is stable.
For $|S|<|S_{\rm c}|$, instead, an instability develops and the system moves towards the SUSY vacuum.

Using the $U(1)$ R symmetry we can rotate $S$ on the real axis.
Then, we define a real scalar field $\sigma$ 
\begin{equation}
S=\frac{\sigma}{\sqrt{2}}
\end{equation}
which is  almost canonically normalized  provided that its value remains
 well below unity in size. In terms of $\sigma$ the potential with $\Phi=\bar \Phi=0$
 becomes
\begin{eqnarray}
\label{pot2}
V_F&\overset{\Phi=\bar \Phi=0}=&\kappa^2 M^4
\left( \left(1+\frac{1}{2}\sigma^2+ \frac{\alpha}{8}\sigma^4\right)^2\left(1+\frac{\alpha}{2}\sigma^2\right)^{-1}\right.
\nonumber\\
&&\quad\quad\quad\quad\left.-\frac{3}{2}\sigma^2\right)\exp\left({\frac{1}{2}\sigma^2+\frac{\alpha}{16}\sigma^4}\right).
\end{eqnarray}
If $\sigma$ satisfies the inequalities $1\gg \sigma^2>
\sigma_{\rm c}^2\simeq 2M^2$, the potential in Eq.~(\ref{pot2})
may be approximated by its second-order expansion in $\sigma^2$
\begin{equation}
\label{pot3}
V_F\overset{\Phi=\bar \Phi=0}\simeq \kappa^2M^4\left(1- \frac{\alpha}{2}\sigma^2+
\frac{1}{8}\left(1-\frac{7}{2}\alpha+2\alpha^2\right)\sigma^4\right)
\end{equation}
which is dominated by the constant term and leads to a (hybrid) inflationary stage.

To the inflationary potential we add the contribution
\begin{equation}
V_{\rm {rad}}= \kappa^2M^4\left(\frac{\delta_{\phi}}{2}\right)
\ln \frac{\sigma^2}{\sigma_{\rm c}^2}
\end{equation}
from radiative corrections, where
\begin{equation}
\delta_{\phi}=N_{\phi}\frac{\kappa^2}{8\pi^2}
\label{deltah}
\end{equation}
and $N_{\phi}=2$ is the dimensionality of the representation to 
which $\Phi$, $\bar \Phi$ belong. Note that this equation is accurate 
for $\sigma /\sigma_{\rm c}\gg1$, which requires that 
$\kappa$ be not much smaller than about $0.01$. The 
potential during inflation is then taken to be 
\begin{equation}
\label{pinf}
V_{\rm {inf}}= \kappa^2M^4\left(1+\frac{\delta_{\phi}}
{2}\ln \frac{\sigma^2}{\sigma_{\rm c}^2}
- \frac{\beta}{2}\sigma^2+\frac{\beta^2\gamma}{4\delta_{\phi}}\sigma^4\right).
\end{equation}
Here
\begin{equation}
\beta=\alpha
\end{equation}
is the negative mass squared of $\sigma$ in units of the false vacuum energy
density $\kappa^2M^4$ ($i.e., m_{\sigma}^2=-\kappa^2M^4\beta$) and
\begin{equation}
\gamma=\frac{\delta_{\phi}}{2\beta^2}\left(1-\frac{7}{2}\alpha+2\alpha^2\right).
\label{gam}
\end{equation}
We assume that $\alpha$ is sufficiently small ($\alpha \lesssim 0.3596$) such that $\gamma >0$.

The first, second, and third derivative of $V_{\rm inf}$  with 
respect to the inflaton field $\sigma$ are, respectively, given
by
\begin{equation}
\label{slope}
V_{\rm {inf}}^{\prime}=\kappa^2M^4
\left(\frac{\delta_{\phi}}{\sigma}\right)\left(1-x+\gamma x^2\right),
\end{equation}
\begin{equation}
V_{\rm {inf}}^{\prime\prime}=- \kappa^2M^4\left(\frac{\delta_{\phi}}{\sigma^2}\right)
\left(1+x-3\gamma x^2\right),
\end{equation}
and
\begin{equation}
V_{\rm {inf}}^{\prime\prime\prime}=2 \kappa^2M^4
\left(\frac{\delta_{\phi}}{\sigma^3}\right)\left(1+3\gamma x^2\right).
\end{equation}
Here, we have introduced the variable
\begin{equation}
x \equiv \frac{\beta\sigma^2}{\delta_{\phi}}.
\end{equation}
Then, from the above relations using the approximation
$V_{\rm {inf}}\simeq \kappa^2M^4$ we obtain the slow-roll parameters
\begin{equation}
\epsilon\equiv\frac{1}{2}\left(\frac{V_{\rm inf}^{\prime}}
{V_{\rm inf}}\right)^2
\simeq \frac{1}{2} \delta_{\phi}\beta\frac{(1-x+\gamma x^2)^2}{x},
\end{equation}
\begin{equation}
\eta\equiv\frac{V_{\rm {inf}}^{\prime \prime}}{V_{\rm inf}}
\simeq -\beta\frac{1+x-3\gamma x^2}{x}
\end{equation}
and
\begin{equation}
\xi^2\equiv\frac{V_{\rm {inf}}^{\prime}V_{\rm {inf}}^{\prime \prime\prime}}{V^2_{\rm inf}}
\simeq 2\beta^2\frac{1-x+\gamma x^2}{x^2}\left(1+3\gamma x^2\right).
\end{equation}
From these equations, assuming that $x$ and $\gamma$ are not much larger than 1
and $\delta_{\phi} \ll 1$, we obtain $\epsilon \ll |\eta|$ and $\epsilon |\eta| 
\ll \xi^2$.

Inflation ends when $\sigma$ reaches the value 
$\sigma_{\rm {end}}$ with
\begin{equation}
\sigma_{\rm {end}}^2\simeq \max\{2M^2, \delta_{\phi} \},
\end{equation}
depending on whether termination of inflation occurs through 
the waterfall mechanism or because of the radiative 
corrections becoming strong ($|\eta| \simeq 1$).

Let the value of the inflaton field $\sigma$ at horizon exit 
of the pivot scale $k_*=0.05 \ \rm {Mpc}^{-1}$ be $\sigma_*$, 
with $x_*$ being the corresponding value of the parameter $x$. 
 The number 
$ N_*$ 
of e-foldings in the slow-roll approximation for the period 
in which $\sigma$ varies between an initial value 
$\sigma_{*}$ and the final value $\sigma_{\rm {end}}$ 
corresponding, respectively, to the values $x_{*}$ 
and
\begin{equation}
x_{\rm {end}}\simeq\beta \max\{ \frac{2 M^2}{\delta_{\phi}}, 1\}
\label{xend}
\end{equation}
of the variable $x$ is given by
\begin{equation}
\label{ef}
N_*\simeq\frac{1}{2\beta}\int_{x_{\rm {end}}}^{x_*}
\frac{dx}{1-x+\gamma x^2}.
\end{equation}
Moreover, the scalar spectral index $n_{\rm s}$ is obtained 
in terms of $\eta_*$, the slow-roll parameter $\eta$ 
evaluated at the value $x=x_*$, as follows:
\begin{eqnarray}
\label{spec}
n_{\rm s}& \simeq& 1+2\eta_* \nonumber\\
&\simeq& 1-2\beta\frac{1+x_*-3\gamma x_*^2}{x_*}.
\end{eqnarray}
Its running $\alpha_{\rm s}\equiv d n_{\rm s}/d \ln{k}$ is
\begin{equation}
\alpha_{\rm s}\simeq -2{\xi_*}^2 \simeq -4\beta^2 
\frac{1-x_*+\gamma x_*^2}{x_*^2}\left(1+3\gamma x_*^2\right),
\end{equation}
where $\xi_*^2$ is the parameter $\xi^2$ evaluated 
at horizon exit of the pivot scale. The scalar 
potential on the inflationary path can be written in terms 
of the scalar power spectrum amplitude $A_{\rm s}$ and the 
value $\epsilon_*$ of the slow-roll parameter $\epsilon$, 
both evaluated at horizon exit of the pivot scale, as
\begin{equation}
V_{\rm {inf}}=24\pi^2\epsilon_*A_{\rm s},
\end{equation}
from which we obtain
\begin{equation}
M^4\simeq 3\beta\left(\frac{N_{\phi}}{2}\right)
\frac{(1-x_*+\gamma x_*^2)^2}{x_*}A_{\rm s}.
\end{equation}
Finally, the tensor-to-scalar ratio is
\begin{equation}
r=16\epsilon_*\ll 1.
\end{equation}

From  Eq.~(\ref{slope}) it follows that
\begin{equation}
\label{slope1}
\sigma V_{\rm {inf}}^{\prime}=\kappa^2M^4\delta_{\phi}\left(
1-x+\gamma x^2\right).
\end{equation}
Thus, it becomes apparent that $\sigma V_{\rm {inf}}^{\prime}$ is not necessarily
positive for all values of the field $\sigma$ satisfying $\sigma /\sigma_{\rm c}>1$.
If $4\gamma<1$ the polynomial $1-x+\gamma x^2$
is negative for all values of $x$ lying between its two distinct real and positive roots.
The smallest root $x_{-}$ corresponds to a value of $|\sigma|$ where $ V_{\rm {inf}}$
has a local maximum with the largest root $x_{+}$ corresponding to a value of
$|\sigma|$ where $ V_{\rm {inf}}$ has a local minimum.
In this case a viable inflationary scenario may take place for values of the inflaton field
corresponding to values of $x<x_{-}$  with $x_{-}\simeq 1$ for $\gamma \ll 1$.
There is always the danger, however, of the inflaton being trapped
in values corresponding to $x>x_{-}$ and ending up in the local minimum of  $ V_{\rm {inf}}$.

In the present work we intend to investigate the possibility of having a
viable inflationary scenario in which $\sigma V_{\rm {inf}}^{\prime}\ge 0$
holds for all $\sigma /\sigma_{\rm c}>1$.  From Eq.~(\ref{slope1})
$\sigma V_{\rm {inf}}^{\prime}\ge 0$ is achieved  if we assume that
$4\gamma \ge 1$ such that the polynomial $1-x+\gamma x^2$ has either
a double real root or a pair of complex conjugate roots. It still holds but as
a strict inequality $\sigma V_{\rm {inf}}^{\prime}> 0$ 
if in the expansion of the potential we include terms up to $\sigma^{10}$
the coefficients of which are positive provided $\alpha\lesssim 0.9$.
Moreover, it can be shown that the exact expression for
$\sigma V_F^{\prime}$ derived from Eq.~(\ref{pot2}) is positive for
$|\sigma|>\sqrt{2\beta/(1-3\alpha/2)}$. Since, as it turns out, we  will be interested
in $\beta\lesssim 0.01-0.02$ we conclude that the possible extema of the
potential occur for values of $\sigma$ for which the analysis based
on the expansion of the potential during inflation to order $\sigma^{10}$
is reliable.

\begin{figure}[t]
\centerline{\epsfig{file=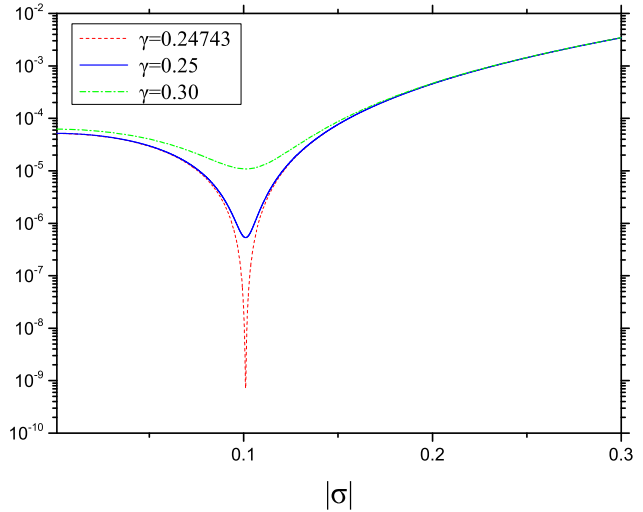,width=11cm}}
\caption{The exact value of $\sigma V_F^{\prime}/(\kappa^2M^4)$ during inflation
with the radiative corrections included as a function of the absolute value of
the field $\sigma$ for $\alpha=0.01$ and $\gamma=0.24743, 0.25, 0.30$. }

\label{figsl}
\end{figure}

In Fig.~\ref{figsl} we plot $\sigma V_F^{\prime}/(\kappa^2M^4)$ using the
exact expression  derived from Eq.~(\ref{pot2}) for $\alpha=0.01$ and
$\gamma=0.24743$,$ 0.25, 0.30$ with the radiative corrections included. For larger
values of $|\sigma|$ than the ones presented in the graph $\sigma V_F^{\prime}$
is certainly positive according to our earlier discussion. We find a minimum at
$|\sigma|\simeq 0.101$ which is very close to the $\gamma$-independent location
\begin{equation}
|\sigma_{\rm m}|=\sqrt{\beta/(1-7\alpha/2+2\alpha^2)}
\end{equation}
of the minimum of the approximate expression given in Eq.~(\ref{slope1}).
The value  of $\sigma V_F^{\prime}$ at the minimum, however, is not zero for
$\gamma=0.25$ but for a slightly smaller value of $\gamma$ which  is close to
$\gamma=0.24743$ for $\alpha=0.01$. As $\alpha$ decreases the vanishing of
$\sigma V_F^{\prime}$ occurs for values of $\gamma$ closer to $\gamma=0.25$.
Thus, it is confirmed that the inflationary potential for $\gamma \ge 0.25$  is strictly
monotonic.

For values of $|\sigma|$ close to $|\sigma_{\rm m}|$, where $\sigma V_F^{\prime}$
is minimized and is close to zero  for $\gamma\simeq 0.25$, the effect of higher
order terms in the expansion of the potential in powers of $\sigma^2$ may not be
negligible. However, this will not affect the ``observable"  inflation if $|\sigma_* |$ is sufficiently
smaller than $|\sigma_{\rm m}|$. We expect that this will be the case because of the
enhanced flatness of the potential for $|\sigma|$ close to $|\sigma_{\rm m}|$.

We first consider the very important special case $\gamma=1/4$ which is amenable
to simple analytic treatment. For this value of $\gamma$
\begin{equation} 
1-x+\gamma x^2=\frac{1}{4}(2-x)^2
\end{equation}
and
\begin{equation}
1+x-3\gamma x^2=\frac{1}{4}(2-x)(2+3x).
\end{equation}
Let us also assume, as it can be verified a posteriori, that
\begin{equation}
x_{\rm {end}}\simeq\beta \ll 1.
\end{equation}
Then,
\begin{equation}
N_*\simeq \frac{1}{\beta}\left(\frac{x_*}{2-x_*}\right)-\frac{1}{2}
\end{equation}
and
\begin{eqnarray}
n_{\rm s}&\simeq&1-\beta \left(\frac{2-x_*}{x_*}\right)\left(1+\frac{3}{2}x_*\right) \nonumber\\
&\simeq& 1-\left(1+\frac{3}{2}x_*\right)\left(N_*+\frac{1}{2}\right)^{-1}.
\end{eqnarray}
From the above equations we easily obtain
\begin{equation}
x_*\simeq\frac{2}{3}\left(\left( N_*+\frac{1}{2}\right)\left(1-n_{\rm s}\right)-1\right)
\end{equation}
 and
\begin{equation}
\beta\simeq \left(\frac{x_*}{2-x_*}\right)\left(N_*+\frac{1}{2}\right)^{-1}.
\end{equation}
Thus, we are able to compute $x_*$ and $\beta=\alpha$ with $N_*$ and $n_{\rm s}$ as inputs.
Finally, using the values of $\gamma$, $\beta=\alpha$ and $x_*$ we may derive the value of the
coupling $\kappa$ and the absolute value $|\sigma_*|$  of the inflaton field at horizon exit of the
pivot scale. In particular, it holds that
\begin{equation}
\kappa^2\simeq (4\gamma)\left(\frac{8\pi^2}{N_{\phi}}\right) \frac{\beta^2}{2-7\alpha+4\alpha^2}
\end{equation}
and
\begin{equation}
\sigma_*^2 \simeq (4\gamma)\frac{x_*\beta}{2-7\alpha+4\alpha^2}.
\end{equation}
As $n_{\rm s}$ decreases with $N_*$ kept fixed both $x_*$ and $\beta$ increase
(assuming $x_*<2$) and from the above equations both $\kappa$ and $|\sigma_*|$
also increase provided, of course, that $\beta=\alpha<0.3596<7/8$.

Throughout the subsequent discussion we make the choice $N_*=50$,  in order to solve
the horizon and flatness problems for reheat temperature $T_{\rm r}\sim 10^9~{\rm GeV}$,
and also set $N_{\phi}=2$ and $A_{\rm s}=2.215 \times 10^{-9}$ \cite{planckcosm}.

Then, for $\gamma=1/4$ and $n_{\rm s}=0.96$ we obtain $x_* \simeq 0.68$,
$\beta=\alpha \simeq 0.0102$, $\kappa \simeq 0.0461$, $|\sigma_*| \simeq 0.06$,
$|\sigma_{\rm m}| \simeq 0.1028$, $\alpha_{\rm s}\simeq -5.28 \times 10^{-4}$,
and $M\simeq 2.086 \times 10^{-3}$.
If, instead, $n_{\rm s}=0.97$ we obtain $x_* \simeq 0.3433$, $\beta=\alpha \simeq 0.0041$,
$\kappa \simeq 0.0184$, $|\sigma_*| \simeq 0.0267$, $|\sigma_{\rm m}| \simeq 0.0645$,
$\alpha_{\rm s}\simeq -4.27 \times 10^{-4}$, and $M\simeq 2.473 \times 10^{-3}$.
Finally, we may also consider the value $n_{\rm s}=0.95$ although at present it does not seem to be
favored by the cosmological data. It gives $x_* \simeq 1.0167$, $\beta=\alpha \simeq 0.02047$,
$\kappa \simeq 0.0944$, $|\sigma_*| \simeq 0.1058$, $|\sigma_{\rm m}| \simeq 0.1485$,
 $\alpha_{\rm s}\simeq -6.96 \times 10^{-4}$, and $M\simeq 1.672 \times 10^{-3}$.

Let us now turn to the more general case where $\gamma>1/4$. From Eq.~(\ref{spec})
we obtain
\begin{equation}
\label{xstar}
x_*\simeq \frac{n_{\rm s}+2\beta-1+\sqrt{\left(1-n_{\rm s}-2\beta\right)^2
+48\beta^2\gamma}}{12\beta\gamma}.
\end{equation}
Also from Eq.~(\ref{ef}), assuming again that $ x_{\rm {end}}\simeq\beta$, we get
\begin{equation}
\label{Nstar}
N_*\simeq \frac{1}{\beta\sqrt{4\gamma-1}} \left( \arctan\frac{1-2 \beta\gamma}
{\sqrt{4\gamma-1}}- \arctan\frac{1-2 x_*\gamma}{\sqrt{4\gamma-1}}\right).
\end{equation}
If the value of $x_*$ from Eq.~(\ref{xstar}) is used in Eq.~(\ref{Nstar}) we obtain $N_*$ as
a function of $\beta$ for a given value of $n_{\rm s}$. Thus, the value of $\beta$ can be determined
(numerically) as the one which gives the desirable value of $N_*$. In the event that there are more
than one such values we choose the smallest one because it leads to the smallest value of
$|\sigma_*|$. The determination of the remaining parameters proceeds as in the previous case.

\begin{figure}[t]
\centerline{\epsfig{file=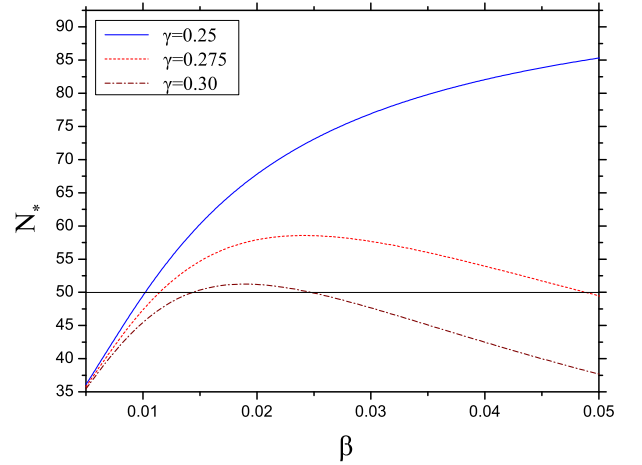,width=10cm}}
\caption{The number $N_*$ of e-foldings as a function of $\beta$ with
$n_{\rm s}=0.96$ for $\gamma=$ $0.25, 0.275, 0.30$. }

\label{figef1}
\end{figure}

\begin{figure}[t]
\centerline{\epsfig{file=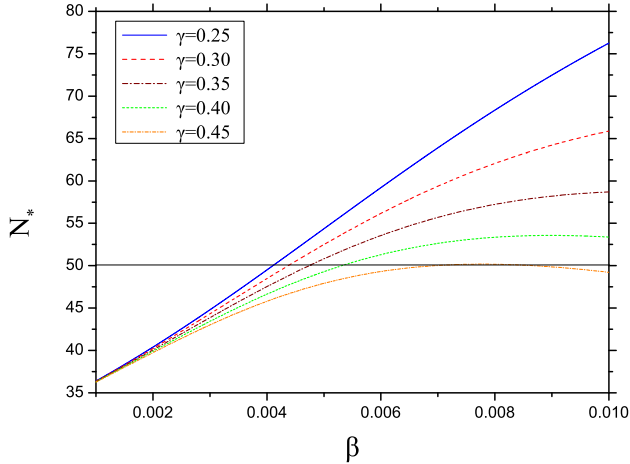,width=10cm}}
\caption{The number $N_*$ of e-foldings as a function of $\beta$ with
$n_{\rm s}=0.97$ for $\gamma=$ $0.25, 0.30, 0.35, 0.40, 0.45$. }

\label{figef2}
\end{figure}

In  Fig.~\ref{figef1} we plot the number of e-foldings $N_*$ as a function of the parameter $\beta$
for $\gamma=0.25, 0.275, 0.30$ assuming $n_{\rm s}=0.96$. Given our requirement that $N_*=50$,
values of $\gamma \gtrsim 0.30$ should not be considered. The chosen value $N_*=50$ seems to
be attained for each allowed value of $\gamma$ at two different values of $\beta$ out of which
we accept the smallest one. We see that $\beta$ lies in the range $\sim 0.01-0.015$.
Similarly, in  Fig.~\ref{figef2} we plot  $N_*$ as a function of $\beta$ for
$\gamma=0.25,  0.30, 0.35, 0.40, 0.45$ assuming $n_{\rm s}=0.97$. We see that
with $n_{\rm s}$ increasing the allowed range of values of $\gamma$ becomes larger while the
range of the parameter $\beta$ is shifted towards smaller values. For  $n_{\rm s}=0.97$ the range
of $\beta$ is $\sim 0.004-0.007$.

Let us consider the case where $\gamma=0.30$, a value allowed  for all values of $n_{\rm s}$
in the range $0.96-0.97$.
If $n_{\rm s}=0.96$ we obtain $x_* \simeq 0.8587$, $\beta=\alpha \simeq 0.01437$,
$\kappa \simeq 0.0718$, $\alpha_{\rm s}\simeq -6.75 \times 10^{-4}$,
$|\sigma_*| \simeq 0.0883$, $|\sigma_{\rm m}| \simeq 0.1231$,
and $M\simeq 1.955 \times 10^{-3}$. 
If, instead,  $n_{\rm s}=0.97$ we get $x_* \simeq 0.3628$, $\beta=\alpha \simeq 0.00437$,
$\kappa \simeq 0.0214$, $\alpha_{\rm s}\simeq -4.4 \times 10^{-4}$,
$|\sigma_*| \simeq 0.0311$, $|\sigma_{\rm m}| \simeq 0.0666$,
and $M\simeq 2.461 \times 10^{-3}$. 

We also give results for $\gamma=0.45$ which is close to the largest allowed value of $\gamma$
for  $n_{\rm s}=0.97$. We have $x_* \simeq 0.5323$, $\beta=\alpha \simeq 0.00694$,
$\kappa \simeq 0.0419$, $\alpha_{\rm s}\simeq -5.6 \times 10^{-4}$,
$|\sigma_*| \simeq 0.0584$, $|\sigma_{\rm m}| \simeq 0.0843$,
and $M\simeq 2.354 \times 10^{-3}$.

One conclusion that can be drawn from Figs.~\ref{figef1} and \ref{figef2} and is confirmed by the
numerical results presented above is that given $n_{\rm s}$ the inflationary scenario with
monotonic $V_{\rm {inf}}$ having the lowest value of $\beta=\alpha$,
the smallest coupling $\kappa$, and the smallest inflaton field value $|\sigma_*|$
is obtained for the smallest value of $\gamma$, namely $\gamma=1/4$. This demonstrates the
 importance of this special case for which it just so happens that it can be treated fully analytically.
 
\section{The initial conditions}
\label{ini}

The above discussion of the hybrid inflationary scenario is certainly simplified
since it is restricted to field values along the inflationary trajectory $\Phi=\bar \Phi=0$.
The naturalness of the scenario, however, depends on the existence of 
field values which although initially are far from the inflationary trajectory
they approach it during the subsequent evolution.
We assume that the energy density 
$\rho$ of the universe is dominated by the F-term potential $V_F$ in Eq.~(\ref{pot1}).
To evade the steep region of the potential generated by supergravity
we require that all field values are well below unity in magnitude such that we are allowed 
to neglect, to a first approximation, terms of dimension higher than four in $V_F$.
Let us start away from the inflationary trajectory and choose the initial energy
density $\rho _{0}$ to satisfy the relation $\kappa^2M^4\ll \rho _{0}\lesssim 1.$
Moreover, we assume that $|\Phi|, |\bar \Phi|$ start somewhat below $|S|$.
Then, $V_F \sim \kappa^2|S|^2\left(|\Phi|^2+|\bar \Phi|^2\right)$.
We would like  $\Phi, \bar \Phi$ to oscillate from the beginning as 
massive fields due to their coupling to $S$ and quickly approach zero.
In contrast, $|S|$ should stay considerably larger than
$|S _{c}|\simeq M$. Thus, initially it must certainly hold that
${m_{ S }^2}\lesssim V_F\lesssim {m_{\Phi ,{\bar \Phi}}^2}$ or
$|\Phi|^2+ |\bar \Phi|^2 \lesssim 1 \lesssim |S|^2$ which contradicts our
assumption that $|S| < 1$. Then we are left with the choice
$\rho_{0} \sim \kappa^2M^4$. In this case $|S|$ remains larger than $|S_{\rm c}|$
provided ${m_{S }^2}\lesssim {\kappa^2M^4}$ or $|\Phi| ^2+|\bar \Phi|^2
\lesssim M^4$. We see that we are forced to start very close to the inflationary
trajectory and severely fine tune the starting field configuration \cite{tet}.

This severe fine tuning becomes more disturbing since the
field configuration at the assumed onset of inflation
should be homogeneous over dinstances $\sim H_{\rm {inf}}^{-1}$,
where $H_{\rm {inf}}$ is the Hubble parameter $H$ at the onset of inflation.
Homogeneity over a Hubble distance, however, is a justified assumption
only if it concerns the end of the Planck era  ($\rho _0\simeq 1$)
where initial conditions should be set. Homogeneity over distances $\sim H_{\rm {inf}}^{-1}$
at the assumed onset of inflation is a natural consequence of  homogeneity over distances
$\sim {H_0}^{-1}$ at the end of the Planck era if during the intervening period
the universe expands by at least a factor $ H_{\rm {inf}}^{-1}/{H_0}^{-1}$ which
corresponds to a minimum  required number
\begin{equation}
\label{reqef}
N_{\rm {req}}=\frac{1}{2} \ln \frac{\rho _0}{ \rho_{\rm {inf}}}
\end{equation}
of e-foldings of expansion of the scale factor $R$.
According to the expansion law $R\sim \rho ^{-\frac{1}{3\gamma_{\rm e} }}$,
however, the number of e-foldings is
\begin{equation}
\label{acef}
N_{\rm {\gamma_{\rm e}}}=\frac{1}{3\gamma_{\rm e} } \ln  \frac{\rho _0}{ \rho_{\rm {inf}}}.
\end{equation}
Typically, $\gamma_{\rm e} \gtrsim 1$ and consequently $N_{\rm {req}} >N_{\rm {\gamma_{\rm e}}}$.
This means that the initial field configuration must be very homogeneous over $ d_{\rm {hom}}$
 Hubble lengths with
\begin{equation}
d_{\rm {hom}} \sim e^{N_{\rm {req}} -N_{\rm {\gamma_{\rm e}}}} =
\left(\frac{\rho _0}{ \rho_{\rm {inf}}}\right) ^
{\frac{3\gamma_{\rm e} -2}{6\gamma_{\rm e} }}\gg 1.
\end{equation}
Such a homogeneity is hard to understand unless an early  period of inflation took
place at $\rho \sim 1$ \cite{initial, pana1, pana2}  with a number of e-foldings
\begin{equation}
\label{early}
N_{\rm {early}}\ge \frac{3\gamma_{\rm e} -2}{6\gamma_{\rm e} }
\ln \frac{\rho _0}{ \rho_{\rm {inf}}} +\frac{1}{3\gamma_{\rm e} }
\ln \frac{\rho_1}{\rho_2}.
\end{equation}
Here $\rho_1=\rho_{\rm {beg}}$ and $\rho_2=\rho_{\rm {end}}$ where 
$\rho_{\rm {beg}}$ ($\rho_{\rm {end}}$) is the energy density at the beginning (end)
of the early inflation. Notice that in Eq.~(\ref{early}) $N_{\rm {early}}$ could be taken to be 
the number of e-foldings of expansion for any period during which $\rho$ varies from
$\rho_1$ to $\rho_2$ with $\rho_0 \ge \rho_1 \ge \rho_{\rm {beg}}$ and
$\rho_{\rm {end}} \ge \rho_2 \ge \rho_{\rm {inf}}$. Setting $\rho_1=\rho_0$
and $\rho_2=\rho_{\rm {inf}}$ the right-hand-side (r.h.s) of Eq.~(\ref{early}) becomes
$N_{\rm {req}}$. An early inflationary stage might also eliminate the requirement
of severe fine tuning of the field configuration at $\rho=\rho_0$ since,
in addition to the homogenization of space, it could alter the dynamics
of the evolution of the universe during the period prior to (the later) inflation.

An inflation taking place at energy density $\rho_{\rm {early}}\gg\rho_{\rm {inf}}$,
however, although eliminates existing inhomogeneities it generates new ones due to
quantum fluctuations. Requiring that the gradient energy density resulting from  these
fluctuations not to exceed $\rho_{\rm {inf}}$ when $\rho \sim \rho_{\rm {inf}}$
gives an upper bound on the energy density $\rho_{\rm {early}}$ (towards the end)
of the first stage of inflation \cite{pana2}
\begin{equation}
  \label{bound1}
\rho _{\rm {early}}\lesssim \left( 6\pi \right) ^{\frac{3\gamma_{\rm e} }{3\gamma_{\rm e} -1}}
\rho_{\rm {inf}} ^{\frac{3\gamma_{\rm e} -2}{2(3\gamma_{\rm e} -1)}} \qquad
\left(\gamma_{\rm e} \gtrsim 1 \right)
\end{equation}
which is somewhat lower than unity and decreases with $\rho_{\rm {inf}}$.

For the role of the early inflation we are going to employ the ``chaotic" D-term inflation
of \cite{pana1} \cite{pana2}. Let $Z$ be a $G_{\rm{LR}}$-singlet chiral superfield with
charge $-1$ under an ``anomalous" $U(1)$ gauge symmetry. The D-term associated with it is
\begin{equation}
V_D=\frac{g^{2}}{2}\left( {K_Z}Z-\xi \right) ^{2}
\end{equation}
with $K_Z$ denoting the derivative of the  K\"ahler potential with respect to $Z$.
If during some period of time $\left|{K_Z}{Z}\right|\ll\xi$ the D-term potential becomes
approximately constant
\begin{equation}
V_D \simeq \frac{1}{2}g^{2}\xi ^{2}
\end{equation}
and on the condition that this constant dominates the energy density the universe experiences
a period of quasi-exponential expansion. In the standard D-term inflation $\left|{K_Z}{Z}\right|$
is kept small because the scalar field $Z$ finds itself lying close to zero trapped in a wrong vacuum.
In the ``chaotic" D-term inflation, instead, $Z$ is not trapped in a wrong vacuum and its initial value
does not have to be small. The version of ``chaotic" D-term inflation that we consider here  and
which we are going to review briefly relies on a rather specific value of the the Fayet-Iliopoulos
$\xi$ term \cite{pana2}.

Let us consider the double-well potential
\begin{equation}
\label{dpot}
V_D=\frac{g^2}{2}\left(\frac{\zeta^2}{2} -\xi\right)^2
\end{equation}
involving the real scalar field $\zeta$ with canonically normalized kinetic term.
This is the ``anomalous" D-term potential of a field $Z$ with a minimal K\"ahler potential
which is brought to the real axis ($Z=\mathsf{Re} Z=\frac{\zeta}{\sqrt{2}}$) by a gauge
transformation under the ``anomalous" $U(1)$ .
For $|\zeta| \gg 1$, as well-known, the equation of motion
\begin{equation}
\label{em1}
\frac{dv}{d\zeta}+\sqrt{3\left(\frac{1}{2}v^2+V\right)}+\frac{V^{\prime}}{v}=0
\end{equation}
with 
\begin{equation}
v \equiv \dot \zeta \equiv \frac{d\zeta}{dt}
\end{equation}
admits the approximate inflationary slow-roll solution
\begin{equation}
v=-\sqrt{\frac{2}{3}}g\zeta.
\end{equation}
For the specific value
\begin{equation}
\xi=\frac{1}{3},
\end{equation}
however, the energy $E=\frac{1}{2}v^2+V$ calculated for the above solution
becomes a ``perfect square" and the approximate slow-roll solution becomes exact.
Integration of this exact solution then leads to
\begin{equation}
\label{exact}
\zeta=\zeta_0 e^{-\sqrt{\frac{2}{3}}gt}
\end{equation}
which demonstrates that $\zeta$ does not oscillate but vanishes only asymptotically with time.
Moreover, the condition for inflationary expansion $-{\dot H}/{H^2}=3{E_k}/{E} < 1$,
where $E_k=v^2/2$ and $H=\sqrt{{E}/{3}}$, is violated only for $|\zeta|$
in the interval [$r_{-}, r_{+}$] with $r_{\pm}=\sqrt{2}\pm 2/\sqrt{3}$. 

Starting from a relatively large $\left|\zeta\right|$ the evolution of the field $\zeta$
approaches the solution in Eq.~(\ref{exact}) giving rise to a ``chaotic" inflationary expansion.
After a short break of the inflationary expansion near the minimum of the potential a new
inflationary expansion begins as $\zeta$ approaches the origin. This approach, however,
is combined with a gradual departure from the non-oscillatory solution.
Eventually, $\zeta$ will either stop before reaching the origin or cross the origin with a small
speed. Then, a new inflationary expansion begins as $\zeta$ moves away from the origin.
The duration of the inflationary stages at $\left|\zeta\right| \ll 1$ will, of course, depend on
the accuracy with which the evolution of the field $\zeta$ follows the special solution 
 in Eq.~(\ref{exact}) which in turn depends on the duration of the inflationary stage at
$\left|\zeta\right| \gg 1$.  

The above discussion concentrates on the ``anomalous" D-term which is assumed to be
dominant during the initial stages of the evolution. This assumption certainly places constraints
on the size of the initial values of all fields present in the model including the $Z$ field itself
since they are all involved in the F-term potential. A noticeable contribution of $Z$ to
the F-term potential is the exponential factor $e^{\frac{\zeta^2}{2}}$.
As a consequence of the constraints on the initial value of $\zeta$ only a very short inflation
 at $\left|\zeta\right| \gg 1$ is allowed which necessitates the additional short inflationary
stage at $\left|\zeta\right| \ll 1$ to complement the required expansion and solve the
problem of the initial conditions.

To minimize the involvement of the $G_{\rm{LR}}$-singlet $Z$ in the F-term potential
we assume that it does not enter the superpotential at all because of the
 charge assignments of the remaining fields under the ``anomalous" $U(1)$ gauge
symmetry. In particular, $S, \Phi, \bar \Phi$ are singlets under the ``anomalous" $U(1)$.
Moreover, we supplement the  K\"{a}hler potential in Eq.~(\ref{kaehler})
with the term
\begin{equation}
\delta K=|Z|^2-\xi.
\end{equation}
Then, the F-term potential becomes
\begin{equation}
\label{pot3}
V_F=\kappa^2\left(V_0+|M^2-\Phi \bar \Phi|^2|Z|^2|S|^2\right)\exp(K+\delta K).
\end{equation}
Minimization of the potential  with respect to $Z$ at fixed $S$ satisfying
$1\gg |S|>|S_{\rm c}|\simeq M$ and with $\Phi=\bar{\Phi}=0$ (which is again a stable choice)
essentially amounts to minimizing the ``anomalous" D-term provided that
\begin{equation}
\frac{1}{2} g^2\xi^2 \gg \kappa^2 M^4.
\end{equation}
This gives $|Z|^2=\xi$ and leads to the F-term (hybrid) inflationary potential which is
again approximated as in Eq.~(\ref{pinf}) but now with
\begin{equation}
\beta=\alpha-\xi
\end{equation}
and
\begin{equation}
\label{gamma1}
\gamma=\frac{\delta_{\phi}}{2\beta^2}\left(1-\frac{7}{2}\alpha+2\alpha^2+2\xi \right).
\end{equation}
In the various scenarios concerning the ``observable" inflation with $\gamma$, $N_*$
and $n_{\rm s}$ taken as inputs, the values of $\beta$, $x_*$, $\alpha_{\rm s}$ and $M$
remain the same, $\alpha$ is determined in terms of $\beta$ as $\alpha=\beta+\xi$,
and $\kappa$ and consequently $|\sigma_*|$ increase slightly due the modified relation
defining $\gamma$ given in Eq.~(\ref{gamma1}). This slight change of $\kappa$
affects only the value of the slow-roll parameter $\epsilon$ which, however, does not
lead to a detectable modification of the inflationary scenario since $\epsilon$ remains tiny.

The study of the initial conditions leading to the ``observable" inflation necessitates departure
from the inflationary trajectory which in turns means that we are no longer allowed to set
$\Phi=\bar\Phi=0$. Using the $U_{\rm {B-L}}$ gauge symmetry we can rotate $\Phi$
to the real axis with $\bar\Phi$ remaining, in general, complex. Thus, we may set
\begin{equation}
\Phi=\frac{\varphi}{\sqrt{2}}, \quad \bar\Phi=\frac{\bar\varphi_1+i\bar\varphi_2}{\sqrt{2}},
\end{equation}
where $\varphi, \bar\varphi_1, \bar\varphi_2$ are canonically normalized real scalar fields.

In order to keep the quantum fluctuations generated during the early inflationary stage
under control we choose the initial energy density $\rho_0 \sim 0.1$, somewhat smaller than 1.
This is achieved by setting the coupling $g$ of the ``anomalous"  $U(1)$ gauge symmetry to the
rather small value $g=0.04$ and choosing a moderately large initial value for the field $\zeta$.
The D-term involving the fields $\Phi, \bar \Phi$ is taken to be
\begin{equation}
V_d=\frac{{g^\prime}^2}{2}\left(|\Phi|^2-|\bar \Phi|^2\right)^2
\end{equation}
with $g^\prime=0.7$.
 
Due to its complexity the problem of the initial conditions will be treated only numerically.
We solve the coupled system of differential equations describing the evolution of the real
scalar fields $\zeta, \sigma, \varphi, {\bar \varphi_1}, {\bar \varphi_2}$ with potential
the complete F-term potential $V_F$ in Eq.~(\ref{pot3}) with the addition of the D-term
potentials $V_D$, $V_d$ and the potential $V_{\rm {rad}}$ involving the radiative corrections.
We also take account of the fact that $\sigma$ is not canonically normalized. As an independent
parameter in all graphs we use the number of e-foldings of expansion
\begin{equation}
N_{\rm {exp}} \equiv \int_0^t Hdt
\end{equation}
with $H$ being the Hubble parameter and $t$  the cosmic time starting from the point where
the initial conditions are set. Throughout our discussion the initial  time derivatives of all fields
are assumed to be equal to zero.

\begin{figure}[t]
\centerline{\epsfig{file=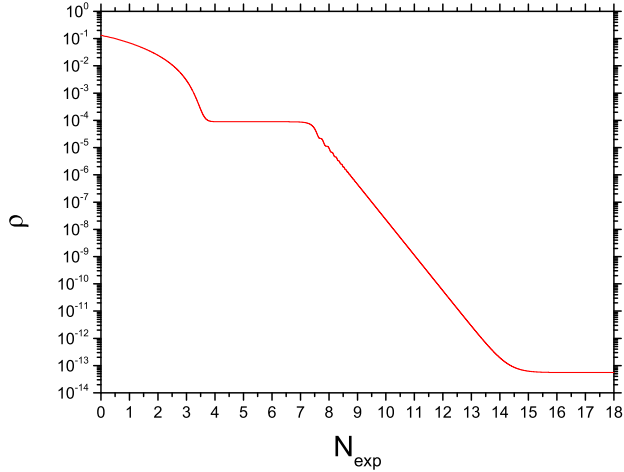,width=10cm}}
\caption{The evolution of the energy density $\rho$ as a function of $N_{\rm {exp}}$
 for $n_{\rm s}=0.96$ and $\gamma=$ $0.25$ with an early inflationary stage.}

\label{figin1}
\end{figure}

\begin{figure}[t]
\centerline{\epsfig{file=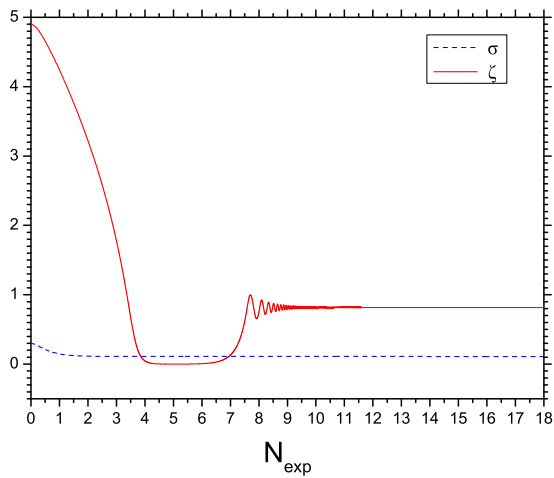,width=10cm}}
\caption{The evolution of the fields $\sigma$ and $\zeta$ as functions of $N_{\rm {exp}}$ 
for $n_{\rm s}=0.96$ and $\gamma=$ $0.25$ with an early inflationary stage.}

\label{figin2}
\end{figure}

\begin{figure}[t]
\centerline{\epsfig{file=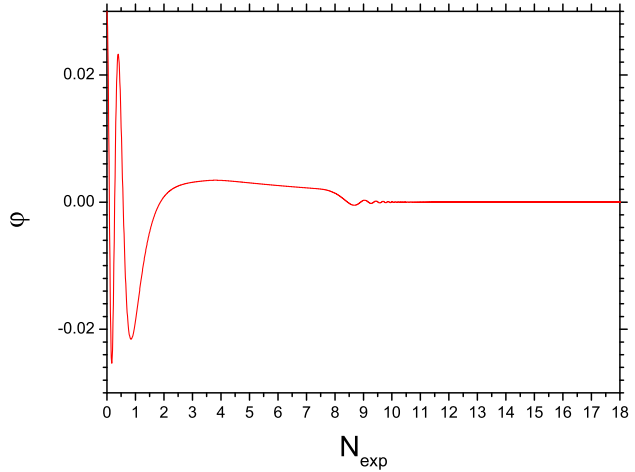,width=10cm}}
\caption{The evolution of the field $\varphi$ as a function of $N_{\rm {exp}}$ 
for $n_{\rm s}=0.96$ and $\gamma=$ $0.25$ with an early inflationary stage.}

\label{figin3}
\end{figure}

We start by considering a choice of initial conditions which lead to the scenario of ``observable"
inflation with  $n_{\rm s}=0.96$ and $\gamma=0.25$. The slightly modified such scenario due
to the presence of the $\xi$ term has $\kappa \simeq 0.0542$ and $\sigma_* \simeq 0.0704$.
The initial field values are chosen to be 
$\zeta=4.9, \sigma=0.3, \varphi=\bar \varphi_1=\bar \varphi_2=0.03$ resulting in an initial
energy density $\rho_0 \simeq 0.1311$. In Fig.~\ref{figin1} we plot the energy density $\rho$,
in  Fig.~\ref{figin2} the values of the fields $\sigma$ and $\zeta$, and finally in  Fig.~\ref{figin3}
the values of the field $\varphi$ as functions of $N_{\rm {exp}}$. For the period from
$N_{\rm {exp}}=0$ to $N_{\rm {exp}}\simeq 2.5$ the system experiences a period of chaotic
inflation with variable energy density during which the value of $\zeta$ varies from $\zeta=4.9$ to
$\zeta \simeq 2.6$. Then, there is a break of the inflationary expansion until $N_{\rm {exp}}\simeq 3.7$
($\zeta\simeq 0.26$) and a new inflation begins at almost constant energy density
$\rho \simeq g^2\xi^2/2 \simeq 8.89 \times 10^{-5}$ during which $\zeta$ first approaches zero and
then moves away from it. This lasts until $N_{\rm {exp}}\simeq 7.4$ ($\zeta \simeq 0.45$).
Soon afterwards $\zeta$ moves towards its minimum at $\zeta=\sqrt{2 \xi}\simeq 0.8165$ and oscillates
around it. This era of matter domination with $\gamma_{\rm e}\simeq 1$  covers the period until the
onset of the later inflation at $\rho \simeq 3\kappa^2M^4 \simeq 1.67 \times 10^{-13}$ and
$N_{\rm {exp}}\simeq 14.07$. From that point on $\gamma_{\rm e} < 2/3$.

The r.h.s. of  Eq.~(\ref{early}) with $\rho_0=\rho_1\simeq 0.1311$,
$\rho_{\rm {inf}}\simeq 1.67 \times 10^{-13}$, $\rho_2=10^{-6}$, and  $\gamma_{\rm e}= 1$
acquires approximately the value $8.493$ while $N_{\rm {exp}}\simeq 8.739$ at $\rho=\rho_2$.
We see that with  $N_{\rm {early}}=N_{\rm {exp}}\simeq 8.739$ the inequality
in  Eq.~(\ref{early}) is satisfied. If instead we take
$\rho_2=\rho_{\rm {inf}}\simeq 1.67 \times 10^{-13}$ again with $\rho_0=\rho_1\simeq 0.1311$
the r.h.s of  Eq.~(\ref{early}) becomes $N_{\rm {req}}\simeq 13.695$
and  $N_{\rm {early}}$ equals $N_{\rm {exp}}\simeq 14.07$ at $\rho=\rho_2$.
We see that in this case the inequality in  Eq.~(\ref{early}) is more amply satisfied because
as $\rho$ approaches $\rho_{\rm {inf}}$ the value of the parameter  $\gamma_{\rm e}$
decreases gradually from $1$ to $2/3$.
We conclude that the early inflation is able to provide the necessary homogenization required
in order to allow the onset of the later inflation.

The field $\sigma$ drops to about $1/2$ its initial value during the first e-folding of expansion, to
about $1/3$ during the second e-folding and it remains essentially frozen to a value well above
$\sigma_*$ until the later inflation begins. The field $\varphi$ oscillates rapidly during the first
stage of the early inflation with decreasing amplitude then remains more or less frozen during
the second stage of the early inflation and starts oscillating again with decreasing amplitude
after the end of the early inflation. Anologous behavior exhibit the fields
$\bar \varphi_1, \bar \varphi_2$.

We also provide initial conditions which lead to other scenarios  of ``observable"
inflation with monotonic $V_{\rm {inf}}$.
 For  $n_{\rm s}=0.95$ and $\gamma=0.25$
we have $\kappa \simeq 0.1104$, $\sigma_* \simeq 0.1238$ and initial field values
$\zeta=4.54, \sigma=0.4, \varphi=\bar \varphi_1=\bar \varphi_2=0.03$.
For  $n_{\rm s}=0.96$ and $\gamma=0.30$
we have $\kappa \simeq 0.0841$, $\sigma_* \simeq 0.1035$ and initial field values
$\zeta=4.75, \sigma=0.3, \varphi=\bar \varphi_1=\bar \varphi_2=0.026$.
 For $n_{\rm s}=0.97$ and $\gamma=0.25$ we have $\kappa \simeq 0.0216$,
$\sigma_* \simeq 0.0314$ and initial field values 
$\zeta=4.95, \sigma=0.3, \varphi=0.08, \bar \varphi_1=\bar \varphi_2=0.07$.
Finally, for $n_{\rm s}=0.97$ and $\gamma=0.45$ we have $\kappa \simeq 0.0492$,
$\sigma_* \simeq 0.0686$ and initial field values 
$\zeta=4.85, \sigma=0.3, \varphi=\bar \varphi_1=\bar \varphi_2=0.037$.

In the case of a non-monotonic $V_{\rm {inf}}$ the choice of appropriate initial conditions
becomes much more tricky because we must ensure that  when $\rho \simeq \kappa^2M^4$
it holds that $|\sigma_*|<|\sigma| <|\sigma_{\rm{lmax}}|$ with $|\sigma_{\rm {lmax}}|$
being the value of $|\sigma|$ for which $V_{\rm {inf}}$ is a local maximum. In addition,
when $\sigma=\sigma_*$ the time derivative of $\sigma$ should be close to the one
predicted by the slow-roll approximation. It is obvious that the outcome is extremely
sensitive to the initial field values. Moreover, we should be aware of the fact that
quantum fluctuations during the early inflationary stage, which are not taken into account
by our classical treatment, could play a crucial role.

\begin{figure}[t]
\centerline{\epsfig{file=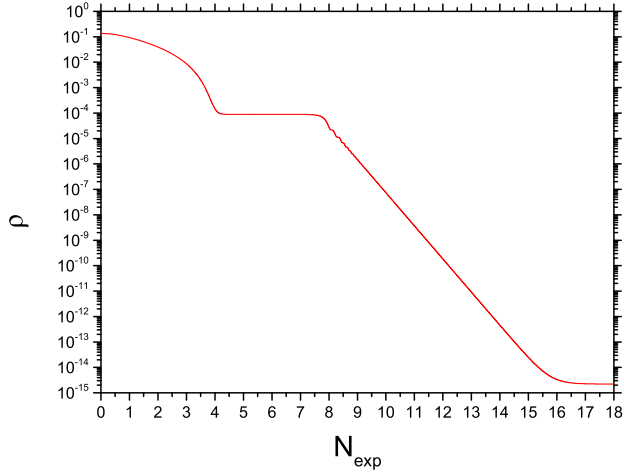,width=10cm}}
\caption{The evolution of the energy density $\rho$ as a function of $N_{\rm {exp}}$
 in the scenario with $\beta=1/150$, $x_*=0.5$ and $\kappa=0.01$
(non-monotonic $V_{\rm {inf}}$) with an early inflationary stage.}

\label{figinit4}
\end{figure}

\begin{figure}[t]
\centerline{\epsfig{file=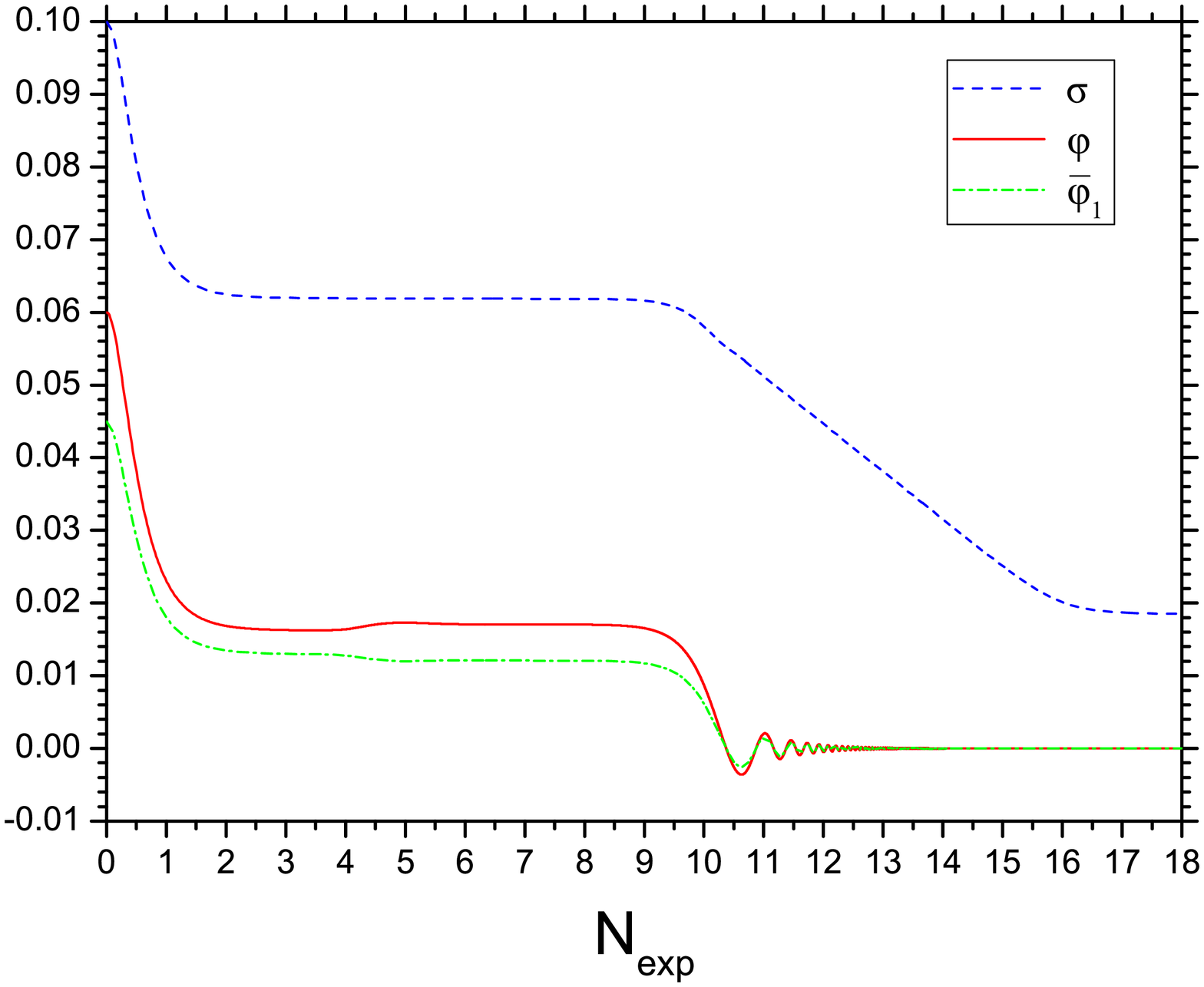,width=10cm}}
\caption{The evolution of the fields $\sigma$, $\varphi$, and $\bar\varphi_1$ as functions
of $N_{\rm {exp}}$  in the scenario with $\beta=1/150$, $x_*=0.5$ and $\kappa=0.01$
(non-monotonic $V_{\rm {inf}}$) with an early inflationary stage.}

\label{figinit5}
\end{figure}

Let us consider the scenario with  non-monotonic $V_{\rm {inf}}$ having
$\beta=1/150$, $x_*=0.5$ and $\kappa=0.01$ \cite{lps}. For such a choice
$n_{\rm s} \simeq 0.96$, $N_* \simeq 50$, $|\sigma_*| \simeq 0.01378$,
$M \simeq 2.17 \times 10^{-3}$, and $|\sigma_{\rm{lmax}}| \simeq 0.0197$.
The initial field values are chosen to be 
$\zeta=5.15, \sigma=0.1, \varphi=0.06, \bar \varphi_1=\bar \varphi_2=0.045$ resulting in an
initial energy density $\rho_0 \simeq 0.1347$. In Fig.~\ref{figinit4} we plot the energy density
$\rho$ and in  Fig.~\ref{figinit5} the values of the fields $\sigma, \varphi$, and $\bar \varphi_1$
as functions of $N_{\rm {exp}}$. The required number of e-foldings of expansion in Eq.~(\ref{reqef})
with  $\rho_0 \simeq 0.1347$ and $\rho_{\rm {inf}}=3\kappa^2M^4 \simeq 6.65 \times 10^{-15}$
is $N_{\rm {req}}\simeq 15.32$ while the actual number is $N_{\rm {exp}}\simeq 15.54$.
We conclude again that the early inflation is able to provide the necessary homogenization required
in order to allow the onset of the later inflation.
The fields $\sigma$, $\varphi$,  $ \bar \varphi_1$
(and also $\bar \varphi_2$) decrease sharply during the first e-folding and remain frozen until
 $N_{\rm {exp}} \simeq 10$.
Then, $\varphi$,  $ \bar\varphi_1$ (and also $ \bar\varphi_2$)
perform damped oscillations with their amplitudes  becoming soon very small while $\sigma$ suffers
a gradual decrease in magnitude until the later inflation begins. When
$\rho \simeq \kappa^2M^4 \simeq 2.22 \times 10^{-15}$ $\sigma$ is clearly below
 $\sigma_{\rm {lmax}} \simeq 0.0197$ and well above  $\sigma_* \simeq 0.01378$ as required.

Comparing the cases of monotonic and non-monotonic inflationary potential we see that in the latter
the initial ratio $|\sigma/\varphi|$ is considerably smaller. In the case of the non-monotonic
potential, however, if the size of the initial value of the field $\varphi$ (and $ \bar\varphi_1$,
$ \bar\varphi_2$) decreases somewhat, when the energy density falls to
$\rho \simeq \kappa^2M^4 \simeq 2.22 \times 10^{-15}$ $\sigma$ remains well above
$\sigma_{\rm {lmax}} \simeq 0.0197$ and eventually is trapped in the local minimum of the potential.

Our earlier discussion should have made clear that the problem of the initial conditions for inflation
consists of two logically distinct components, namely the initial field values and the creation of a
homogeneous region in space of appropriate size where the fields take these values. If we assume
that this region already exists then it is possible to solve very simply the other part of the initial
condition problem using the same  ``anomalous" D-term potential of Eq.~(\ref{dpot}) in which the
$\xi$-term does not necessarily take the very specific value $\xi=1/3$ and the initial value of
$|\zeta|$ is neither very large nor very small. We choose $\xi=0.5$ and $g=1$ in order to
obtain initial energy density $\rho_0 \sim 0.1$.

\begin{figure}[t]
\centerline{\epsfig{file=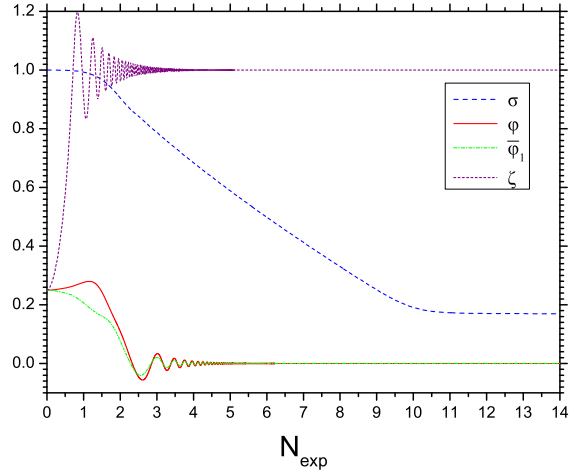,width=10cm}}
\caption{The evolution of the fields $\sigma$, $\varphi$, $\bar \varphi_1$, and $\zeta$ as functions of 
$N_{\rm {exp}}$ for $n_{\rm s}=0.96$ and $\gamma=0.25$ without an early inflationary stage.}

\label{figinit6}
\end{figure}

\begin{figure}[t]
\centerline{\epsfig{file=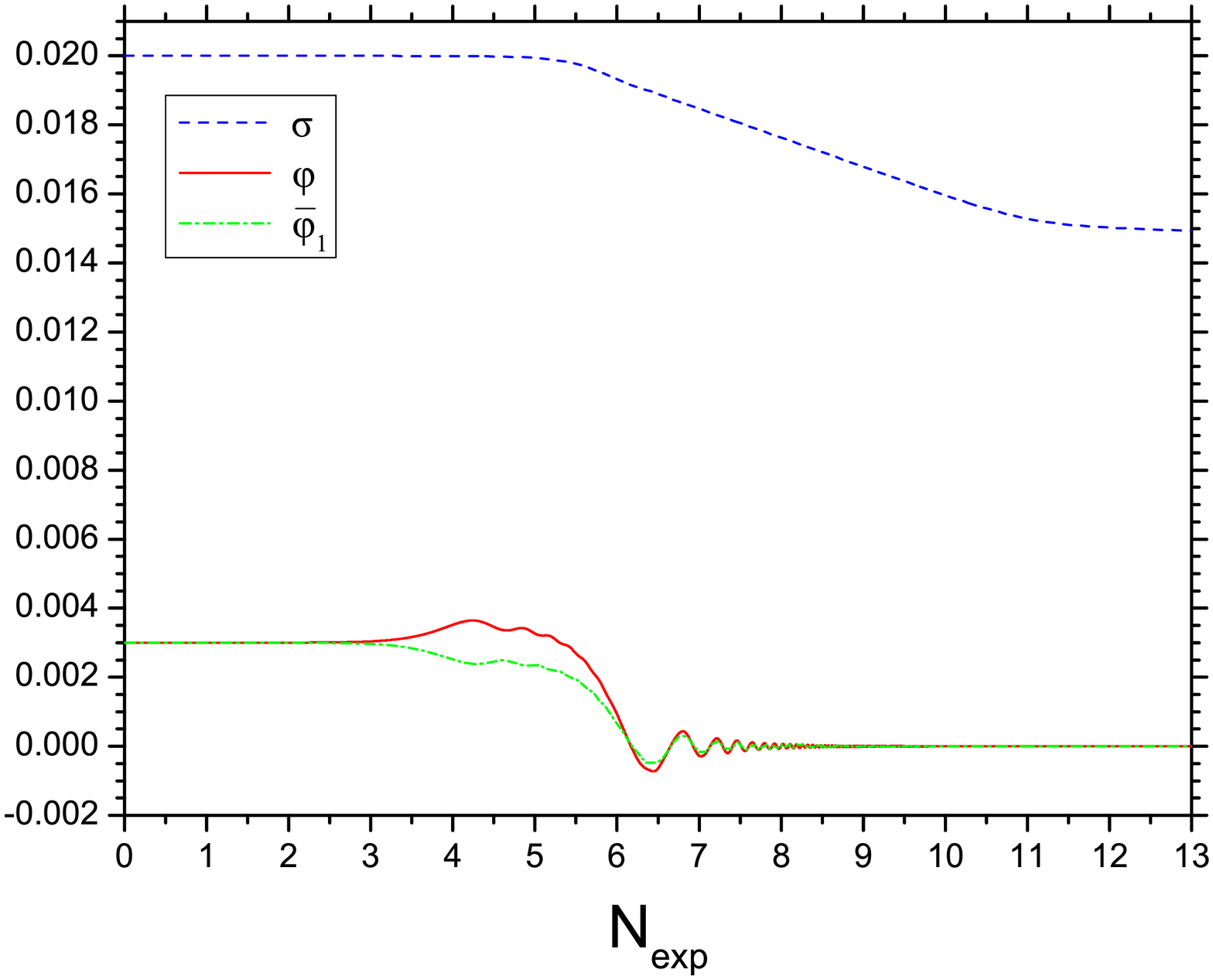,width=10cm}}
\caption{The evolution of the fields $\sigma$, $\varphi$, and $\bar \varphi_1$ as functions of 
$N_{\rm {exp}}$  in the scenario with $\beta=1/150$, $x_*=0.5$ and $\kappa=0.01$
(non-monotonic $V_{\rm {inf}}$) without an early inflationary stage.}
\label{figinit7}
\end{figure}

As a demonstration let us consider the scenario of ``observable" inflation with  $n_{\rm s}=0.96$
and $\gamma=0.25$. Initially we set $\sigma=1$ and $\zeta=\varphi=\bar\varphi_1=\bar\varphi_2=0.25$
such that $\rho_0\simeq 0.1103$. In Fig. \ref{figinit6} we plot the evolution of the fields  $\sigma$,
$\varphi$, $\bar \varphi_1$, and $\zeta$ as functions of $N_{\rm {exp}}$. We see that $\zeta$
starts oscillating fast with decreasing amplitude around its minimum at $\zeta=\sqrt{2\xi}=1$  already
from the first e-folding of expansion. Also $\varphi, \bar \varphi_1$ (and $\bar \varphi_2$) after the
second e-folding perform fast dumped oscillations around zero and soon become very small in size.
Finally, $\sigma$ after the first e-folding decreases continuously until the onset of inflation at 
$N_{\rm {exp}}\simeq 9.6$ reaching a value $\sigma \simeq 0.17 $ considerably larger than
$\sigma_*\simeq 0.07$.

We may also consider the scenario with non-monotonic $V_{\rm {inf}}$ having $\beta=1/150$,
$x_*=0.5$ and $\kappa=0.01$. In this case we choose an initial value of $\sigma=0.02$ very close
to the position of the local maximum of $V_{\rm {inf}}$ in order to minimize the sensitivity to the initial
conditions. The initial values for the remaining fields are $\zeta=0.25$,
and $\varphi=\bar\varphi_1=\bar\varphi_2=0.003 \gg M^2$ leading to $\rho_0\simeq 0.1103$.
The evolution of the fields  $\sigma$, $\varphi$, and $\bar \varphi_1$ as functions of $N_{\rm {exp}}$
 is presented in Fig. \ref{figinit7}. We see that although the initial value of $\sigma$ is not much larger
than $\sigma_*\simeq 0.01378$,  $\sigma$ remains larger than $\sigma_*$ when  inflation begins
($N_{\rm {exp}}\simeq 10.6$).

\section{Summary}
\label{concl}

We investigated the possibility of having a viable scenario of SUSY hybrid inflation with scalar spectral
index $n_{\rm s}\simeq 0.96 - 0.97$ and monotonic inflationary potential. This is achieved
for values of the superpotential coupling $\kappa$ and the coefficient $\alpha$ of the first correction
to the minimal K\"{a}hler potential involving the inflaton for which the quantity $\gamma$ in Eq.~(\ref{gam})
lies in a certain interval. The lower endpoint of this interval is close to $0.25$ with the upper endpoint
being an increasing function of $n_{\rm s}$.

We also provided a solution to the problem of the initial conditions leading to such inflationary
scenarios which employs an additional early inflationary stage. This approach seems to be
applicable to the case of a non-monotonic inflationary potential as well in the sense that it
generates the necessary homogeneous region at the Planck scale where the fields are almost
constant  with values which are not unnaturally small in size. However, the extreme sensitivity to
the choice of these initial field values justifies our preference for monotonic potentials.

\def\plb#1#2#3{{Phys. Lett. B }{\bf #1},~#3~(#2)}
\def\prl#1#2#3{{Phys. Rev. Lett.}
{\bf #1},~#3~(#2)}
\def\prd#1#2#3{{Phys. Rev. D }{\bf #1},~#3~(#2)}
\def\aap#1#2#3{{Astron. Astrophys.}
{\bf #1},~#3~(#2)}

\end{document}